\documentclass{acm_proc_article-sp}
\usepackage{amsmath}
\usepackage{dsfont}
\usepackage{algorithm}
\usepackage{algorithmic}
\usepackage{array}

\newcommand{\out}{\mathrm{out}}
\newcommand{\inn}{\mathrm{in}}
\newcommand{\remove}[1]{}

\begin{document}

\title{Using Proximity to Predict Activity in Social Networks}

\numberofauthors{2} 
%
\author{
%
%
\alignauthor
Kristina Lerman\\
       \affaddr{Information Sciences Institute}\\
       \affaddr{University of Southern California}\\
       \affaddr{Marina del Rey, CA 90292}\\
       \email{lerman@isi.edu}
\alignauthor
Suradej Intagorn, Jeon-Hyung Kang and Rumi Ghosh\\
       \affaddr{University of Southern California}\\
       \affaddr{Los Angeles, CA }\\
       \email{\{intagorn,jeonhyuk,rumig\}@usc.edu}
}
\remove{
\alignauthor Rumi Ghosh\\
       \affaddr{Computer Science Department}\\
       \affaddr{University of Southern California}\\
       \affaddr{Los Angeles, CA }\\
      \email{}
\and  
\alignauthor \\
       \affaddr{Computer Science Department}\\
       \affaddr{University of Southern California}\\
       \affaddr{Los Angeles, CA }\\
       \email{}
}
\date{}

\maketitle
\begin{abstract}
The structure of a social network contains information useful for predicting its evolution. Nodes that are ``close'' in some sense are more likely to become linked in the future than more distant nodes. We show that structural information can also help predict node activity.
We use proximity to capture the degree to which two nodes are ``close'' to each other in the network. In addition to standard proximity metrics used in the link prediction task, such as  neighborhood overlap, we introduce new metrics that model different types of interactions that can occur between network nodes. We argue that the ``closer'' nodes are in a social network, the more similar will be their activity. We study this claim using data about URL  recommendation on social media sites Digg and Twitter. We show that structural proximity of two users in the follower graph is related to similarity of their activity, i.e., how many URLs they both recommend. We also show that given friends' activity, knowing their proximity to the user can help better predict which URLs the user will recommend.  We compare the performance of different proximity metrics on the activity prediction task and find that some metrics lead to substantial performance improvements.
\end{abstract}




\section{Introduction}
The structure of complex networks contains valuable information that can be used to identify missing links and predict which new links between existing nodes are likely to be observed in the near future~\cite{Liben-Nowell03,Koren06,Tong07,linkprediction}. Given a pair of unconnected nodes, link prediction algorithm calculates a graph-based proximity score between them. The ``closer'' the two nodes are, the more likely they are to become linked in the future, or in the case of partially observed networks, the more likely a link to actually exist between them.
Researchers proposed a large variety of proximity metrics for the link prediction task, including local measures, such as the number of common neighbors, the fraction of common neighbors, metrics that weigh the contribution of each common neighbor by the inverse of its degree (linear)~\cite{Zhou09} or the logarithm of its degree (Adamic-Adar)~\cite{adamic03friends}, as well as  global metrics based on the number of paths between nodes (Katz)~\cite{Katz53} or the probability that a random walk starting at one node will reach the other~\cite{Koren06}. A number of studies tested the performance of these metrics on the link prediction task in different networks. Liben-Nowell and Kleinberg~\cite{Liben-Nowell03} showed that Adamic-Adar score best predicts new links in scientific co-authorship networks, with Katz score a close second. Zhou et al.~\cite{Zhou09}, on the other hand, found that the linear version of the Adamic-Adar score best predicts missing links in biological and technological networks, including protein-protein interaction networks, electrical power grid and US air transportation networks. Neither study motivated the metrics or explained how to choose a appropriate metric  for the problem.

Structural proximity measures how readily information can be exchanged by nodes in a network even in the absence of a direct link between them. The greater the number of paths connecting two nodes through intermediaries, the greater the potential for information exchange; therefore, the closer the nodes are. However, the degree to which information can reach one node from another depends not only on network topology, but also on \emph{the nature of the process by which nodes interact}~\cite{Ghosh11nonconservative}. One-to-one interactions, such as phone calls and Web surfing, can be modeled as a random walk. Therefore, metrics based on the random walk, such as conductance~\cite{Koren06}, are appropriate as a proximity measure. The one-to-many interactions common in online social media are fundamentally different and cannot be modeled as a random walk. In social media, rather than picking a network neighbor to whom to transmit information, users broadcast information to all their neighbors. Broadcast-based interactions are best modeled by an epidemic process, and therefore, require a different measure of proximity.
We propose local proximity metrics that take  into account both the topology of the network and the nature of interactions between nodes. We show how these metrics map to the known metrics used in link prediction.

We show that structural proximity metrics can help predict activity in social networks. We illustrate this claim on the benchmark Southern Women data set. Next, we study in detail URL recommendation activity on social media sites Digg and Twitter. These sites allow users to post URLs to stories on the Web, and other users to recommend them to others by voting for them (on Digg) or retweeting them (on Twitter)~\cite{Lerman10icwsm}. Both sites also allow users to follow the activities of others. When a user tweets a URL (or submits one on Digg), the URL is broadcast to all the user's followers, who may in turn decide to retweet it (or vote for it), thereby broadcasting it to their own followers, and so on.  We investigate how well structural proximity metrics based on the follower graph predict whether the user will vote for or retweet the URL. Note that activity prediction differs from the link prediction problem. In the latter, network structure is used both as the basis for prediction and to evaluate prediction results. In activity prediction, on the other hand, prediction results are evaluated independently of the network structure using evidence from users' voting or retweeting behavior.

This paper makes the following contributions:
\begin{itemize}
\item New structural proximity metrics for directed graphs that take into account the nature of interactions between nodes (Section~\ref{sec:proximity})
\item Definition of the activity prediction task for social networks (Section~\ref{sec:activity})
\item Detailed study of the activity prediction task in social media (Section~\ref{sec:social-media})
\end{itemize}

\section{Interactions and Proximity}
\label{sec:proximity}
We represent a network by a directed, unweighted graph $G = (V,E)$ with $V$ nodes and $E$ edges. The adjacency matrix of the graph is defined as: $A(u,v)= 1$ if $(u,v) \in E$; otherwise, $A(u,v)= 0$.
The set of  out-neighbors of $u$ is $\Gamma_{\out}(u) = \lbrace v \in V \vert (u,v) \in E \rbrace$,   and  the out-degree of $u$ is $d_{\out}(u) = \sum_{v \in V}A(u,v)= |\Gamma_{\out}(u)|$, where $|.|$ denotes the size of the set. Similarly, $\Gamma_{\inn}(u)$ represents the set of in-neighbors of $u$, and $d_{\inn}(u)$ is the in-degree of $u$. The total degree of the node is $d(u)=d_{\out}(u)+d_{\inn}(u)$. In undirected graph, the neighborhood of $u$ consists of nodes that are connected to $u$ and is denoted by $\Gamma(u)$.

\remove{
The number of common neighbors is:
\begin{equation}
CN = \frac{1}{2} \big[|\Delta| + |\Delta^{\prime}|\big].
\end{equation}
\noindent Jaccard coefficient measures the fraction of common neighbors:
\begin{equation}
JC =\frac{1}{2}\Big[\frac{|\Gamma_{\out}(u) \cap \Gamma_{\inn}(v)|}{|\Gamma_{\out}(u) \cup \Gamma_{\inn}(v)|} + \frac{|\Gamma_{\out}(v) \cap \Gamma_{\inn}(u)|}{|\Gamma_{\out}(v) \cup \Gamma_{\inn}(u)|}  \Big].
\end{equation}
\noindent The Adamic-Adar score weighs each common neighbor by the inverse of the log of its degree:
\begin{eqnarray}
AA & = & \frac{1}{2}\Big[\sum_{z \in \Delta}{\frac{1}{\log(d(z))}} +
\sum_{z^{\prime} \in \Delta^{\prime}}{\frac{1}{\log(d(z^{\prime}))}}\Big]. \nonumber
\end{eqnarray}
}

\begin{figure}[htbp]
\begin{center}
\includegraphics[width=0.5\linewidth]{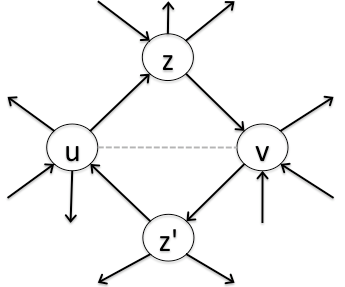}
\caption{Example of a directed graph.}
\label{fig:directed}
\end{center}
\end{figure}

\begin{table}[htdp]
\caption{Some of the proximity metrics used for network analysis, including four proposed in this paper}
\label{metrics}
\begin{center}
\begin{tabular}{|c|c|}
\hline
$metric$ & $definition$ \\ \hline
$CN$ & $CN = \frac{1}{2} \big[|\Delta| + |\Delta^{\prime}|\big]$ \\ \hline
$JC$ & $
$JC$ =\frac{1}{2}\Big[\frac{|\Gamma_{\out}(u) \cap \Gamma_{\inn}(v)|}{|\Gamma_{\out}(u) \cup \Gamma_{\inn}(v)|} + \frac{|\Gamma_{\out}(v) \cap \Gamma_{\inn}(u)|}{|\Gamma_{\out}(v) \cup \Gamma_{\inn}(u)|}  \Big]$
\\ \hline
$AA$ & $AA  =  \frac{1}{2}\Big[\sum_{z \in \Delta}{\frac{1}{\log(d(z))}} + \sum_{z^{\prime} \in \Delta^{\prime}}{\frac{1}{\log(d(z^{\prime}))}}\Big] $ \\ \hline
\hline
$CS$ & $CS = \frac{1}{2} \sum_{z \in \Delta}{\frac{1}{d_{\out}(u) d_{\out}(z)}} $ \\
& $+ \frac{1}{2}\sum_{z \in \Delta^{\prime}}{\frac{1}{d_{\out}(v) d_{\out}(z)}} $ \\ \hline
$CS\_AL$ & $CS\_AL  =  \frac{1}{2}\sum_{z \in \Delta}{\frac{1}{d_{\out}(u) d_{\inn}(z) d_{\out}(z) d_{\inn}(v)}}$  \\
& $+ \frac{1}{2} \sum_{z \in \Delta^{\prime}}{\frac{1}{d_{\out}(v)  d_{\inn}(z) d_{\out}(z)  d_{\inn}(u)}}$ \\ \hline
$NC$ & $NC = \frac{1}{2} \big[|\Delta| + |\Delta^{\prime}|\big] $ \\
& \\ \hline
$NC\_AL$ & $NC\_AL = \frac{1}{2}\sum_{z \in \Delta}{\frac{1}{d_{\inn}(z) d_{\inn}(v)}}$\\
& $+ \frac{1}{2} \sum_{z \in \Delta^{\prime}}{\frac{1}{d_{\inn}(z) d_{\inn}(u)}}$ \\ \hline
\end{tabular}
\end{center}
\label{default}
\end{table}%

Intuitively, network proximity measures the likelihood a message starting at node $u$ will reach $v$, regardless of whether an edge exists between $u$ and $v$. The greater the number of paths connecting $u$ and $v$, the more likely they are to share information, and the closer they are considered to be in the network.
Proximity metrics used in previous studies~\cite{Liben-Nowell03,linkprediction} include the number of common neighbors (CN), fraction of common neighbors, or Jaccard (JC) coefficient, and the Adamic-Adar (AA) score, which weighs each common neighbor by the inverse of the logarithm of its degree. Table~\ref{metrics} gives their definition in terms of the directed neighborhoods of $u$ and $v$:
\begin{eqnarray}
\Delta & = & \Gamma_{\out}(u) \cap \Gamma_{\inn}(v) \nonumber\\
\Delta^{\prime} & = & \Gamma_{\inn}(u) \cap \Gamma_{\out}(v). \nonumber
\end{eqnarray}

The likelihood a message will reach $v$ from $u$ depends, however, not only on the number of paths, but also on the nature of the dynamic process by which messages spread on the network~\cite{Ghosh11nonconservative}. Consider a graph of hyperlinked Web pages. The process of browsing this graph is best described by a random walk. At each page, a Web surfer picks one of the neighbors of that page in the Web graph and navigates to it. The interactions by which information is exchanged in  the air transportation network, the electric power grid and mobile phone network can also be modeled by the random walk. We call such processes \emph{conservative}, since they conserve some underlying mass distribution. Not all interactions, however, are conservative. The one-to-many interactions common in social media, where users broadcast information to all their followers, cannot be modeled as a random walk. This, and many other social phenomena, such as the spread of disease or innovation, are \emph{non-conservative} in nature, since  the amount of information, disease or innovation in the network does not remain constant.
Different dynamic processes will lead to different notions of proximity, even in the same network. In this section, we describe two classes of processes and the proximity metrics they lead to.

\paragraph{Conservative proximity}
Consider conservative processes first. Koren et al.~\cite{Koren06} introduced cycle-free effective conductance as a measure of proximity. This is a global metric that computes the probability a random walk starting at $u$ will reach $v$ through any path in the graph.  In the directed graph shown in Fig.~\ref{fig:directed}, a walker starting at $u$ can reach $v$ through $z$. It is possible that longer paths exist connecting $u$ to $v$, but we do not consider them, since in most cases we are interested in \emph{local} measures, that depend only on the neighborhoods of $u$ and $v$. Such measures  are not only easier to compute, but they also do not require knowledge of the full graph, e.g., the entire Twitter follower graph, which is difficult to obtain. Local proximity will only consider paths between $u$ and $v$ that go through a single intermediate node, e.g., $z$ or $z^{\prime}$ in Fig.~\ref{fig:directed}. To reach $z$ from $u$, the random walker needs to pick the correct edge, which it will do with probability $1/d_{\out}(u)$, and it will reach $v$ from $z$ with probability $1/d_{\out}(z)$. Symmetrizing, we obtain conservative proximity measure, which gives the probability a random walk will reach $u$ from $v$ or vice versa through paths of length two:
\begin{equation}
\label{eq:cs}
CS = \frac{1}{2}\Big[\sum_{z \in \Delta}{\frac{1}{d_{\out}(u) d_{\out}(z)}} + \sum_{z \in \Delta^{\prime}}{\frac{1}{d_{\out}(v) d_{\out}(z)}}\Big].
\end{equation}
\noindent Note that in an undirected graph, this metric reduces to
\begin{equation}
\label{eq:cs-undirected}
\overline{CS} = \frac{1}{2} \Big[\frac{1}{d(u)}+\frac{1}{d(v)}\Big] \sum_{z \in \Gamma(u) \cap \Gamma(v)}{\frac{1}{d(z)}}.
\end{equation}
\noindent Like the Adamic-Adar score, conservative proximity takes into account the degree of the common neighbor. This measure is almost identical to the resource allocation metric (RA) shown by Zhou et al.~\cite{Zhou09} to be the best-performing local metric on the missing link prediction task in several networks, including the network of political blogs, the electric power grid, router-level Internet graph, and US air transportation network.  On an undirected network RA is:
$$
RA=\frac{1}{d_u}\sum_{z \in \Gamma(u) \cap \Gamma(v)}{\frac{1}{d(z)}}.
$$
\noindent Conservative proximity in undirected networks (Eq.~\ref{eq:cs-undirected}) is the symmetric version of this metric. Therefore, RA metric should work well on these networks, because, except for political blogs, the processes taking place on them are conservative in nature. In other words, when a plane leaves one airport, its destination can be exactly one airport.
For the political blogs network, Zhou et al. ignored the direction of links, which may have changed properties of the network.

Social networks, especially online social networks, are composed of actors with a limited resource, their attention~\cite{Wu07}. We model limited attention by forcing nodes to monitor a small number of their in-links at a time. This alters the dynamic process and affects propagation of messages. Now, in order for a message to get from $u$ to $z$, it must not only go over the correct out-link from $u$, but $z$ must also pay attention to that in-link to receive the message, which it will do with probability $1/d_{in}(z)$. Attention limited conservative proximity metric can be written as:
\begin{eqnarray*}
CS\_AL & = & \frac{1}{2}\Big[\sum_{z \in \Delta}{\frac{1}{d_{\out}(u) d_{\inn}(z) d_{\out}(z) d_{\inn}(v)}}  \\
& & + \sum_{z \in \Delta^{\prime}}{\frac{1}{d_{\out}(v)  d_{\inn}(z) d_{\out}(z)  d_{\inn}(u)}}\Big].
\end{eqnarray*}

\paragraph{Non-conservative proximity}
Now imagine that information flows on a network via one-to-many broadcasts. When a node broadcasts a message, it is sent to all the node's out-neighbors. In this case, for a message to get from $u$ to $v$ in Fig.~\ref{fig:directed}, first $u$ broadcasts it to its neighbors, including $z$, and then $z$  broadcasts it. For a message to get from $v$ to $u$, $v$ broadcasts it and then $z^{\prime}$  broadcasts it. Probability of the message being transmitted from one node to another is one. Therefore, symmetrized non-conservative proximity measure is:
\begin{equation}
NC = \frac{1}{2}\Big[\sum_{z \in \Delta}{1} + \sum_{z \in \Delta^{\prime}}{1}\Big] = \frac{1}{2} \big[|\Delta| + |\Delta^{\prime}|\big].
\end{equation}
\noindent The non-conservative metric counts the expected number of times a message is received and is identical to the neighborhood overlap metric $CN$.  While this metric was originally motivated by the intuition that when people have many friends in common, they are more likely to attend the same events and be in the same community, our work shows that it also can be derived from the principles of non-conservative dynamics, of which social interactions are a prime example.

Finite attention can also play a role in non-conservative interactions. When $u$ broadcasts a message, $z$ will receive it only if it pays attention to the channel from $u$. Therefore, symmetric attention-limited non-conservative proximity metric can be written as
$$
NC\_AL = \frac{1}{2}\Big[\sum_{z \in \Delta}{\frac{1}{d_{\inn}(z) d_{\inn}(v)}} + \sum_{z \in \Delta^{\prime}}{\frac{1}{d_{\inn}(z) d_{\inn}(u)}}\Big].
$$
\noindent
In undirected graphs, this reduces to
$$
\overline{NC\_AL} = \frac{1}{2} \Big[\frac{1}{d(u)}+\frac{1}{d(v)}\Big] \sum_{z \in \Gamma(u) \cap \Gamma(v)}{\frac{1}{d(z)}},
$$
\noindent which is identical to  conservative proximity in undirected networks (Eq.~\ref{eq:cs-undirected}).

\section{Proximity and Activity}
\label{sec:activity}
In social networks,  network proximity can be interpreted as social closeness. In his seminal paper Granovetter~\cite{Granovetter73} argued that the strength of a social tie, which specifies the intensity and the depth of interaction between two people, can be estimated from their local network structure. He proposed neighborhood overlap as the metric to quantify tie strength. Subsequently, a large-scale study of a mobile phone network established a correlation between the strength of ties, measured by the frequency and duration of phone calls between two people, and structural proximity, measured by their neighborhood overlap~\cite{Onnela07}. We claim that proximity also has predictive power.  People who are close to each other in a social network are more likely to act in a similar way because they  share the same information, attend the same events, or participate in the same community. Knowing the actions of some people allows us to predict the actions of others who are close to them in the network.

\begin{figure}[tbh]
\begin{center}
\includegraphics[width=0.85\linewidth]{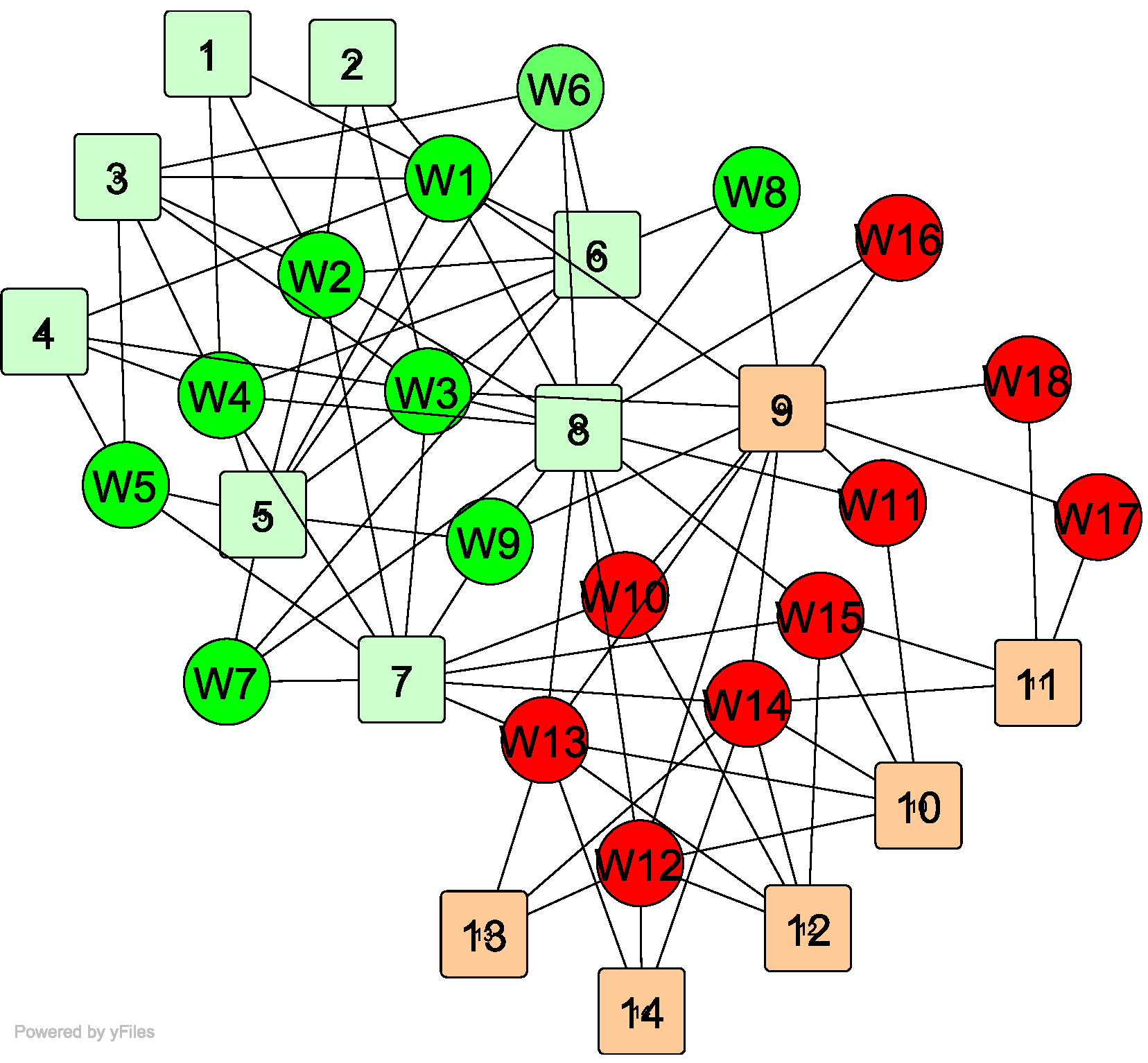}
\end{center}
\caption{Bipartite graph representing the Southern Women dataset.  Circles represent women and squares the events they attended. }
\label{fig:dataset}
\end{figure}

\subsection{Illustration: Southern Women Data Set}
We illustrate activity prediction task on the benchmark Southern Women data set.
This data set  comes from a comparative study of social class by Davis et al.~\cite{DGG}, in which researchers systematically collected  data about the social activities of 18 women over a nine month period. Over this time period, subsets of women met at 14 informal events. Event attendance is shown in Fig.~\ref{fig:dataset}, where circles are women and squares are events. Original researchers identified two groups, or communities, in this network, which later researchers attempted to reconstruct from the network data~\cite{Freeman01}.

\remove{
\begin{figure}[tbh]
\begin{center}
\includegraphics[width=0.85\linewidth]{correlation}
\end{center}
\caption{}
\label{fig:correlation}
\end{figure}
}

\subsubsection{Analysis of Proximity Metrics}
We create a social network of women by projecting the bipartite graph in Fig.~\ref{fig:dataset} onto an unweighted, undirected, unipartite graph, where an edge between two women exists if they attended any event together. We then compute proximity of every pair of women using the metrics defined in Table~\ref{metrics}. The number of events the pair has co-attended quantifies their co-activity, and can also be taken as a measure of tie strength. Proximity values along ties are substantially (16\%--30\%)  higher than for non-linked women. For example, the average number of common neighbors of two women linked by an edge is $13.6$, while for unlinked women it is $10.4$.

\begin{table}[htdp]
\caption{ Correlation between proximity of pairs of women and the number of events they co-attended, for all pairs connected by an edge in the network. }
\label{sw-corr}
\begin{center}
\scalebox{0.9}{
\begin{tabular}{|c|c|c|c|c|c|c|}
\hline
{CN}& {JC} & {AA} & {CS} & {CS\_AL} & {NC} & {NC\_AL} \\
\hline
 0.515 & 0.504 & 0.519 &  \textbf{0.532} &  0.492 & 0.515& \textbf{0.532}\\
 \hline
\end{tabular}
}
\end{center}
\end{table}

Proximity and co-activity are related. The higher the proximity of two women, the greater the number of common events they attended. Table~\ref{sw-corr} reports correlation of proximity and co-attendance along all ties. While all proximity metrics are well correlated with activity, the highest correlation is produced by the conservative (CS) and attention-limited non-conservative (NC\_AL) metrics.

\subsubsection{Predicting Activity}
\label{sec:methodology}
\begin{figure}[tbh]
\begin{center}
\includegraphics[width=0.9\linewidth]{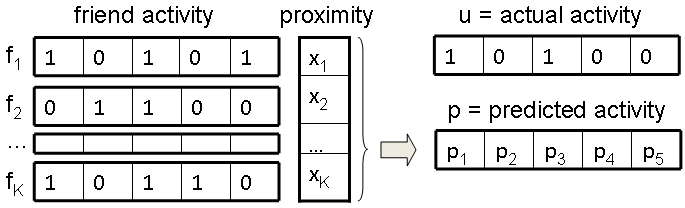}
\end{center}
\caption{Prediction methodology}
\label{fig:methodology}
\end{figure}
Results above suggest that we may use proximity to predict event co-attendance. Specifically, women will attend the same events as their friends, but they are more likely to attend the events that their closer friends attend. Therefore, even if we do not have information about events a woman attended, we may reconstruct it from the events her friends attended.
To quantitatively evaluate this claim, we divide events into a training set, containing $N$ randomly picked events, and a test set, with the remaining $14-N$  events. We construct a unipartite network of women who attended $N$ training events and use this network to compute proximity scores. We represent the test events a woman attended by a binary vector of length $14-N$, whose $i$th value is 1 if the woman attended the $N+i$th event, and 0 otherwise. For each woman, we construct a prediction vector $\vec{p}$ of length $14-N$ that aggregates test events her friends attended, weighing friends by their proximity to the woman, as shown in Fig.~\ref{fig:methodology}. The value $p_i$ of the prediction vector is the weighted number of friends who attended the $N+i$th event. To compute precision and recall of the prediction, we construct a binary vector $\vec{u}$ of test events the woman actually attended. Then precision is $Pr=\vec{u} \cdot \vec{p}/|\vec{p}|$ and recall is $Re=\vec{u} \cdot \vec{p}/|\vec{u}|$, where $|\vec{z}|=\sum_{i}{z_i}$. Algorithm~\ref{alg:predict} gives the pseudo code of the prediction algorithm. As baseline, we create a prediction vector that weighs all friends uniformly, without regard to their actual proximity to the woman.

\begin{algorithm}
\caption{Predict woman $w$'s attendance of test events} 
\label{alg:predict}
\begin{algorithmic}[1]
\STATE $F \Leftarrow$ friends($w$)
\FOR {each friend $j \in F$}
\STATE $\vec{f_j} \Leftarrow$ test\_events($j$) \COMMENT{\emph{vector of test events friend attended}}
\STATE $x_j \Leftarrow$ proximity($w,j$) \COMMENT{\emph{friend's proximity to w}}
\ENDFOR
\STATE $\vec{p} = \sum_j{ \vec{f_j} x_j/|\vec{x}|}$ \COMMENT{\emph{construct prediction vector}}
\STATE $\vec{u} \Leftarrow$ test\_events($w$) \COMMENT{\emph{test events w actually attended}}
\STATE $Pr=\vec{u} \cdot \vec{p}/|\vec{p}|$
\STATE $Re=\vec{u} \cdot \vec{p}/|\vec{u}|$
\end{algorithmic}
\end{algorithm}

\begin{figure}[tbh]
\begin{center}
\begin{tabular}{c} 
\includegraphics[width=0.9\linewidth]{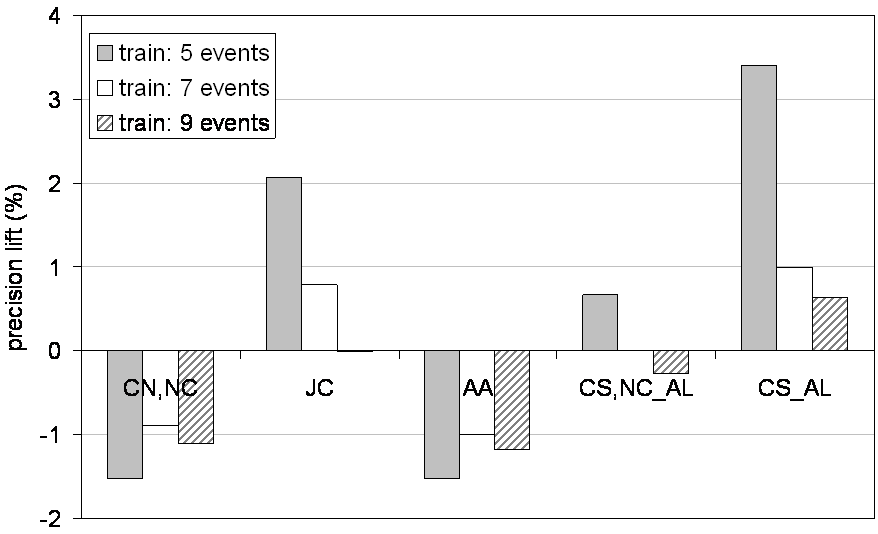}\\
\includegraphics[width=0.9\linewidth]{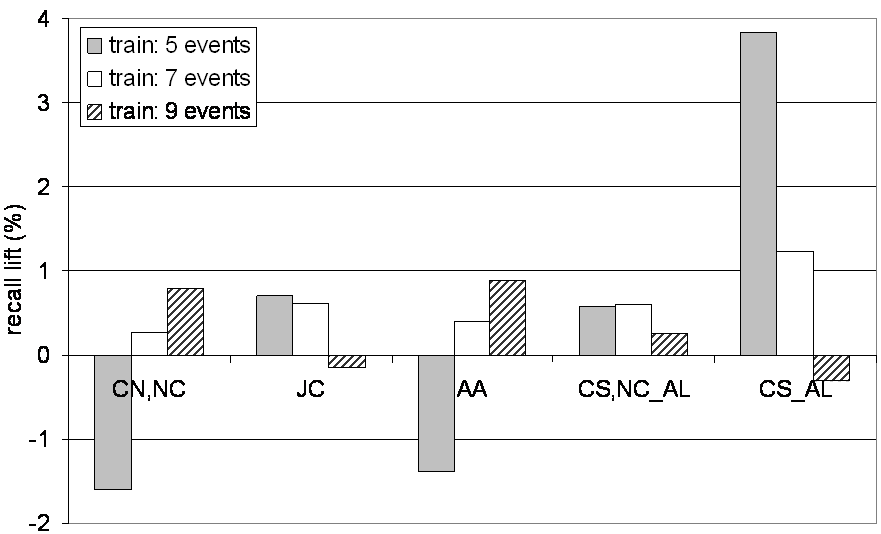}
\end{tabular}
\end{center}
\caption{Precision and recall lift achieved by different proximity metrics on the activity prediction task in the Southern Women data set. Lift is defined as \% change over baseline.}
\label{fig:prediction}
\end{figure}

Figure~\ref{fig:prediction} reports performance of different proximity metrics on the activity prediction task in the Southern Women data set. We used three different training sets: $N=5$, 7, and 9 events. The baseline uniform friend predictor attains precision values in the range 0.52--0.57. Not all proximity-based predictors can beat the baseline: the precision of the common neighbors (CN), non-conservative (NC), and Adamic-Adar (AA) predictions fails to beat baseline in all three experiments. The most lift, i.e., \% improvement over baseline, is attained by conservative, attention-limited conservative,  and Jaccard metrics. Interestingly, we get the most lift on the smallest training set, $N=5$ events. As more data becomes available for proximity, and conversely, less data for prediction, the precision of the best predictors decreases. Similar trends are observed in recall, although while the recall of the best performing metrics decreases with the training set size, the recall of the worst performing metrics increases, and even beats  baseline.

\section{Predicting Activity in Social\\ Media}
\label{sec:social-media}
Social media has emerged as critical platform for disseminating information~\cite{Lerman10icwsm,WuWatts11}, marketing products~\cite{Aral11}, harvesting social knowledge~\cite{Plangprasopchok11wsdm}, and occasionally  stirring political~\cite{Lotan11} and social unrest. While there are many different social media sites which allow for a broad range of activity --- posting updates, sharing photos and videos, tagging content, checking into places --- in this paper we focus on URL recommendation on two popular social media sites: Digg and Twitter. Both sites allow registered users to post URLs to content they find online and other users to recommend these URLs by voting for them on Digg or retweeting them on Twitter. Like many other social media sites, Digg and Twitter allow users to follow the activities of other users by adding them as friends. We call the resulting online social network \emph{follower graph}. Note that the follower graph is directed: when user $A$ adds user $B$ as a friend, $A$ can follow $B$ and see the URLs $B$ recommends, but not \emph{vice versa}, unless $B$ also follows $A$.

When a user recommends a URL, by retweeting or voting for it, she makes it visible to her followers. The followers may in turn vote for or retweet the URL to their own followers, and so on, creating cascades through which information spreads through the follower graph. While several studies have empirically studied diffusion of information in networks~\cite{Lerman10icwsm,www11-hashtags,Versteeg11icwsm}, its  mechanism is hotly debated. Competing theories argue that information spreads because people influence their followers to propagate it, or simply because similar people tend to be linked and exposed to the same information (homophily)~\cite{Anagnostopoulos08,Aral09,Choudhury10}, though the two effects are difficult to tease apart~\cite{Shalizi11}. Rather than contribute to the debate, our goal is to show that information in the follower graph can help predict user activity on these sites. While users tend to recommend URLs their friends recommend, knowing the friends' proximity in the follower graph can help better predict which URLs the user will recommend.

\subsection{Data sets}
\textbf{Digg} (http://digg.com) is a social news aggregator with over 3 million registered users.  Digg allows users to submit links to and recommend news stories by voting on, or {digging}, them. A newly submitted story goes to the {upcoming} stories list, where it remains for 24 hours, or until it is promoted to the {front page} by Digg, whichever comes first. Of the tens of thousands of daily submissions, Digg picks about a hundred to feature on its front page.

We used Digg API to collect complete voting record for all stories promoted to Digg's front page in June 2009.\footnote{http://www.isi.edu/$\sim$lerman/downloads/digg2009.html} The data associated with each story contains story anonymized id, submitter's anonymized id, and list of voters with the time of each vote. We also collected the time each story was promoted to the front page. In total, the data set contains over 3 million votes on 3,553 front page stories.

Of the 139K voters in the data set, more than half followed at least one other user. We retrieved their user names and reconstructed the follower graph of active users. This graph contained 70K nodes and more than 1.7 million edges.

\textbf{Twitter} (http://twitter.com) is a popular social networking site that allows registered users to post and read short text messages (at most 140 characters). A user can also retweet the content of another user's post. Like Digg, Twitter allows users to follow the activity of others.

Twitter's Gardenhose streaming API provides access to a portion of real time user activity, roughly 20\%-30\% of all user activity.\footnote{At present time, Gardenhose is restricted to 10\% of real time content.}
We used this API to collect tweets over a period of three weeks. We focused on tweets that included a URL in the body of the message, usually shortened by some URL shortening service, such as bit.ly or tinyurl. In order to ensure that we had the complete tweeting history of each URL, we used Twitter's search API to retrieve all tweets associated with that URL. Then, for each tweet, we used the REST API to collect friend and follower information for that user. Data collection process resulted in more than 3 million tweets which mentioned 70K distinct shortened URLs.  There were 816K users in our data sample, but we were only able to retrieve follower information for some of them, resulting in a graph with almost 700K nodes and over 36 million edges.

Retweeting activity in our sample encompassed diverse behaviors from spreading newsworthy content to orchestrated human and bot-driven campaigns that included advertising and spam. We recently proposed a novel method to automatically classify these behaviors~\cite{Ghosh11snakdd} by characterizing the dynamics of retweeting with two information theoretic features. The first feature is the entropy of the distinct user distribution, and second feature is the entropy of the distinct time interval distribution. We showed that these two features alone were able to accurately separate activity into meaningful classes. High user entropy implies that many different people retweeted the URL, with most people retweeting it once. High time interval entropy implies presence of many different time scales, which is a characteristic of human activity.  In this paper, we focus  on those URLs from the data set which are characterized by high ($>3$) user and time interval entropies. These parameter values are associated with the spread of news-worthy content and excludes robotic spamming and manipulation campaigns driven by few individuals.
This left us with a data set containing 3,798 distinct URLs retweeted by 542K distinct Twitter users.

\subsection{Analysis of Proximity Metrics}
We compute proximity metrics on the directed follower graphs of active Digg and Twitter users. Proximity metrics used in this study are listed in Table~\ref{metrics}. We measure similarity of activity of a pair of users by the number of common URLs they both recommended. Activity of a pair of Digg users is measured by \emph{co-votes}, the number of promoted stories for which they both voted. Activity of a pair of Twitter users is measured by \emph{co-retweets}, the number of common URLs they both tweeted or retweeted.

\begin{figure*}[htbp]
\begin{center}
\begin{tabular}{|ccc|}
\hline
\multicolumn{3}{|c|}{\textbf{(a) Digg}}\\
\includegraphics[width=0.3\linewidth]{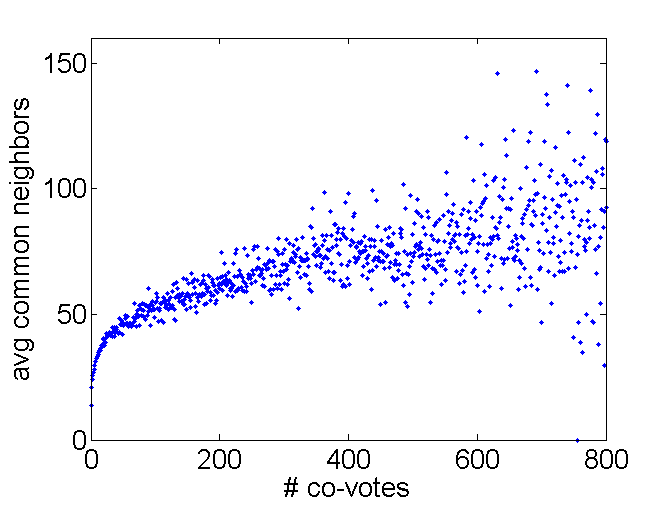} &
\includegraphics[width=0.3\linewidth]{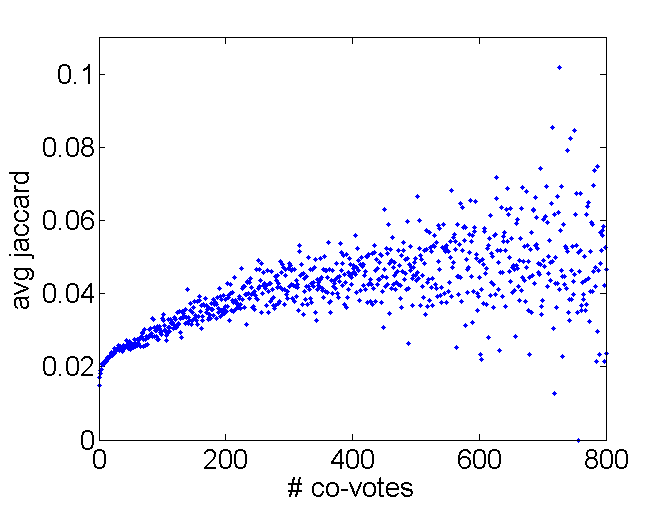} &
\includegraphics[width=0.3\linewidth]{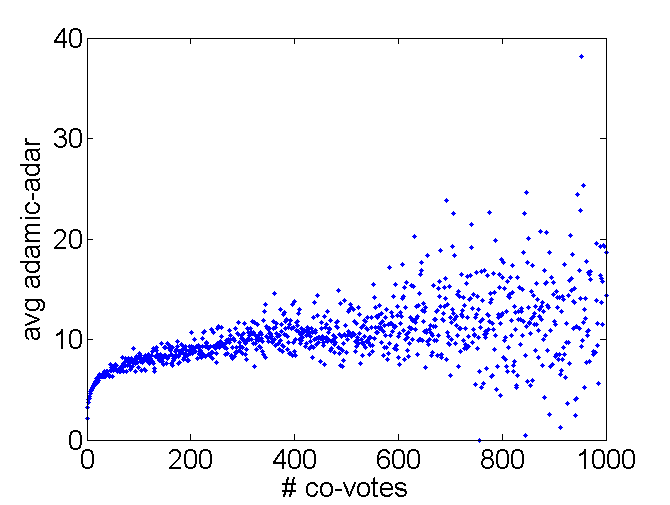} \\
common neighbors (CN, NC) & jaccard (JA) &  adamic-adar (AA)\\
\includegraphics[width=0.3\linewidth]{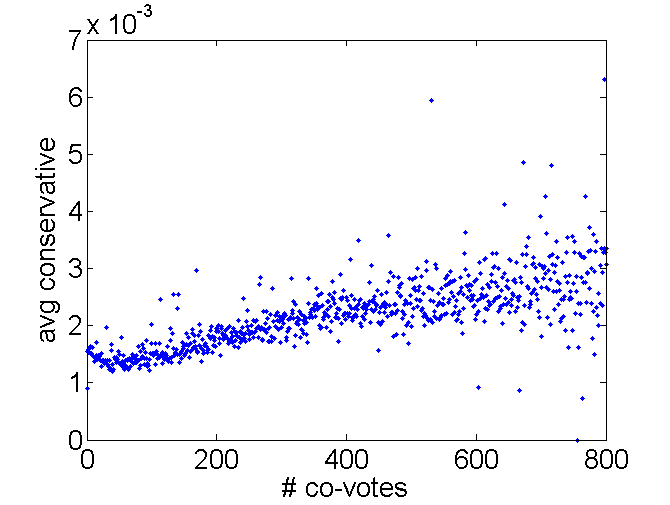} &
\includegraphics[width=0.3\linewidth]{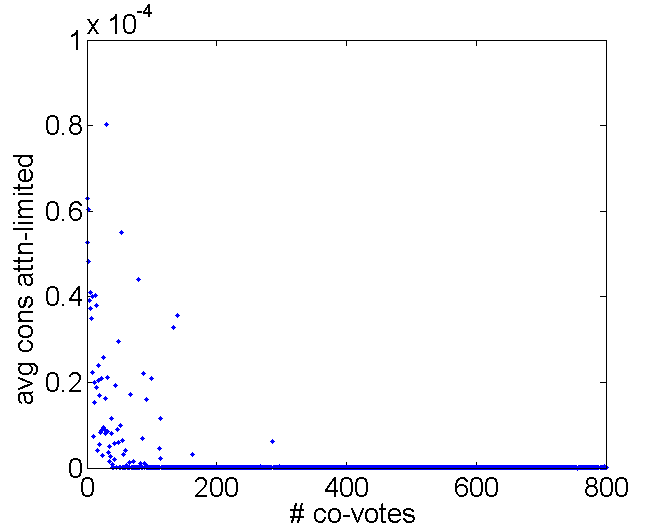} &
\includegraphics[width=0.3\linewidth]{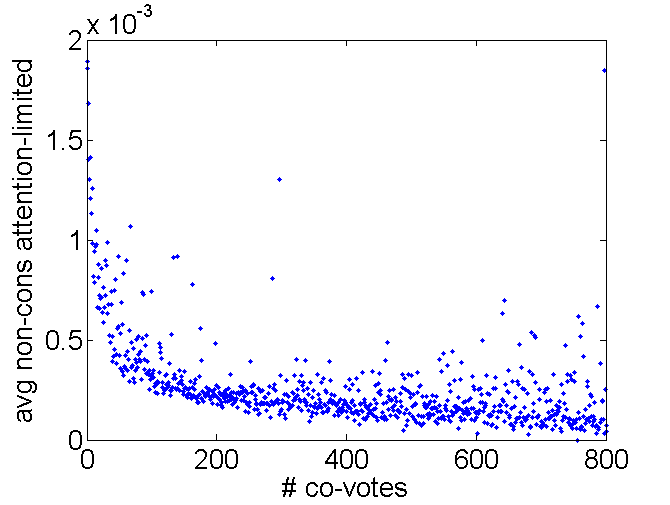} \\
conservative (CS) &  conservative attn-limited (CS\_AL) &  non-cons., attn-limited (NC\_AL)
\\ \hline
\multicolumn{3}{|c|}{\textbf{(b) Twitter}}\\
\includegraphics[width=0.3\linewidth]{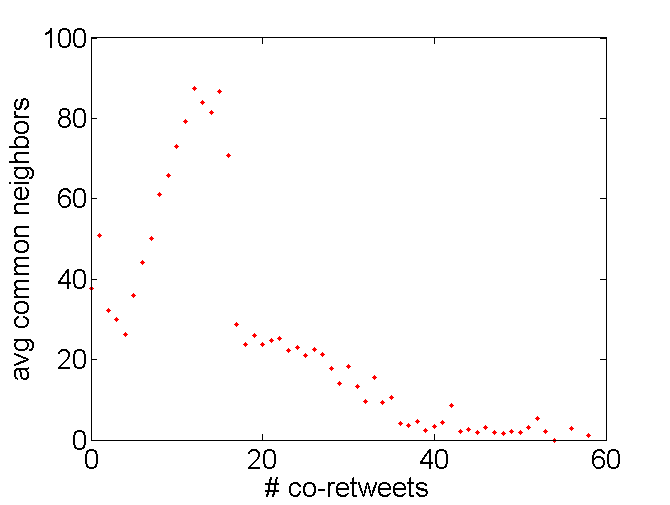} &
\includegraphics[width=0.3\linewidth]{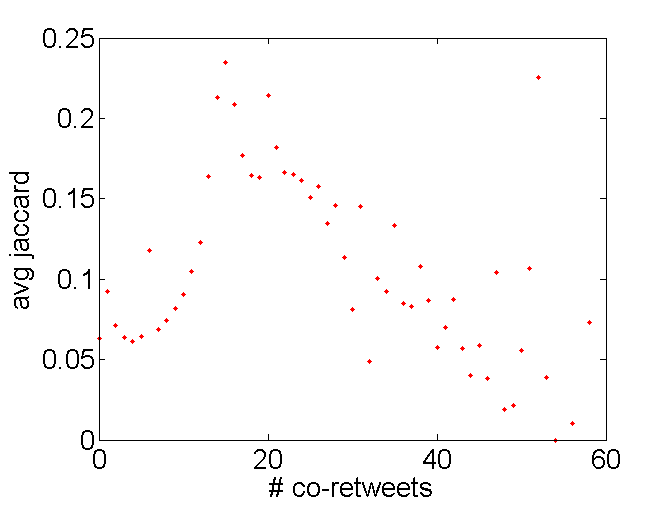} &
\includegraphics[width=0.3\linewidth]{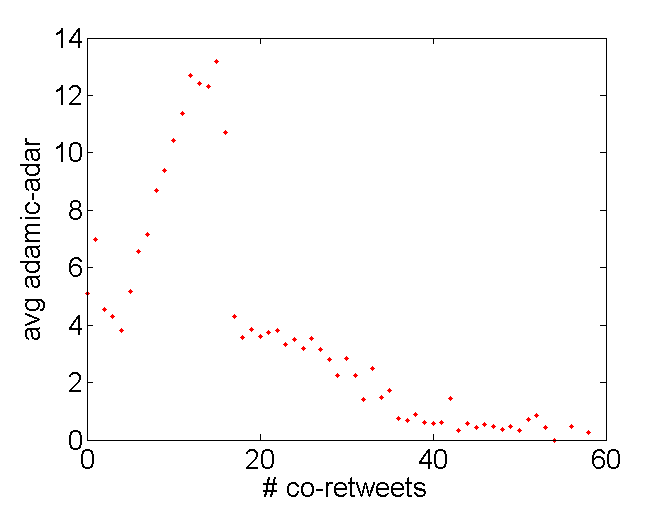} \\
 common neighbors (CN, NC) & jaccard (JA) &  adamic-adar (AA)\\
\includegraphics[width=0.3\linewidth]{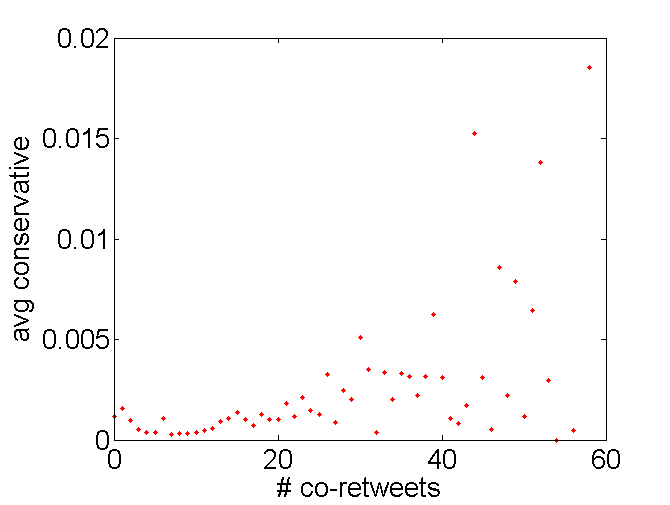} &
\includegraphics[width=0.3\linewidth]{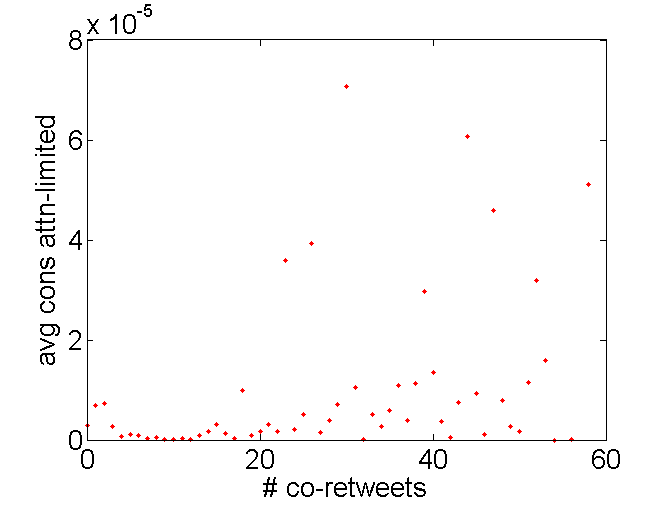} &
\includegraphics[width=0.3\linewidth]{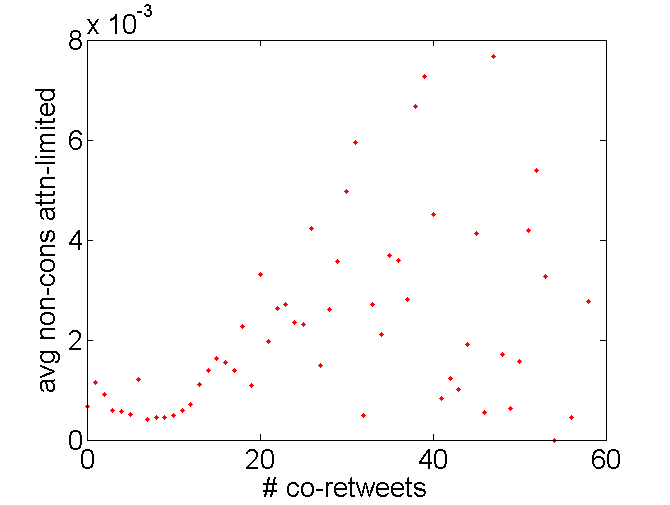} \\
conservative (CS) & conservative attn-limited (CS\_AL) &  non-cons., attn-limited (NC\_AL)
\end{tabular}
\caption{Average value of the proximity metrics vs activity for pairs of users linked by an edge in the follower graphs of (a) Digg and (b) Twitter.}
\label{fig:digg_corr}
\end{center}
\end{figure*}

Figure~\ref{fig:digg_corr} plots proximity, computed using different metrics, vs activity for pairs of users linked by an edge in the follower graph on either site. The y-value represents the average proximity for all pairs with that many co-votes or co-retweets. There are significant trends in proximity as a function of activity on Digg (Fig.~\ref{digg-corr}(a)), at least for co-votes $<800$. Above this value, there is no observable correlation between proximity and activity. This could be because some users tend to vote on many front page stories regardless of their content, or due to automatic voting. Interestingly, attention-limited versions of the conservative and non-conservative proximity decrease with the number of co-votes. Conservative metric is the only one to display a behavior that is not, on the whole, monotonic: the value of the metric decreases until around 50 co-votes and increases after that.

Proximity--activity trends on Twitter are more complex (Figure~\ref{fig:digg_corr}(b)). In the first three plots, the average value of proximity initially increases with activity, until about 15 co-retweets, at which point there is a decreasing trend. The last three metrics, however, show an increasing trend.

\begin{table*}[htbp]
\caption{Correlation between proximity of pairs of users connected by an edge in the follower graph and their co-activity on (a) Digg and (b) Twitter. Rows in (a) present co-votes under different filter conditions. For example, co-votes < 200 condition reports correlations for pairs of users who voted for fewer than 200 common stories. The number of pairs satisfying the filter condition is reported in the second column. }
\begin{center}
\begin{tabular}{|c|c|c|c|c|c|c|c|c|}
\hline
\multicolumn{9}{|c|}{(a) \emph{Digg: correlation}} \\ \hline
filter & \# edges	&	CN	&	JC	&	AA	&	CS	&	CS\_AL	&	NC	&	NC\_AL	\\
\hline
co-votes < 200	&	1,410,590	&	\textbf{0.256}	&	0.129	&	0.232	&	0.015	&	-0.010	&	\textbf{0.256}	&	 -0.028	\\
co-votes < 400	&	1,429,712	&	\textbf{0.277}	&	0.158	&	0.246	&	0.019	&	-0.009	&	\textbf{0.277}	&	 -0.027	\\
co-votes < 800	&	1,438,320	&	\textbf{0.283}	&	0.170	&	0.249	&	0.024	&	-0.008	&	\textbf{0.283}	&	 -0.025	\\
all	&	1,439,842	&	\textbf{0.279}	&	0.163	&	0.246	&	0.025	&	-0.008	&	\textbf{0.279}	&	-0.023	 \\
 \hline
\multicolumn{9}{|c|}{(b) \emph{Twitter: correlation}} \\ \hline
& \# edges & CN	&	JC	&	AA	&	CS	&	CS\_AL	&	NC	&	NC\_AL	\\\hline
& 28M & -0.769 & -0.339 &-0.755 &  \textbf{0.523} & 0.350 & -0.769 & \textbf{0.406} \\ \hline
\end{tabular}
\end{center}
\label{digg-corr}
\end{table*}

We  compute correlation between proximity and activity for all pairs of  users linked by an edge in the follower graph. These correlations for different proximity metrics are shown in Table~\ref{digg-corr}.  We can limit the edges taken into account by correlation to those that satisfy some filter condition. For example, co-votes < 200 line reports correlations for pairs of Digg users who voted on fewer than 200 common stories. The number of pairs satisfying the filter condition is reported in the second column.  Despite growing scatter, correlation increases with the amount of co-activity until about 800 co-votes. The non-conservative metric, which is equivalent to the common neighbors metric, leads to highest correlation. The story is somewhat different for Twitter (Table~\ref{digg-corr}(b)), where the conservative and attention-limited non-conservative metrics lead to highest correlations.

\subsection{Prediction Results}
Social media users tend to act like the people they follow. This means that users tend to vote for stories their friends vote for on Digg~\cite{Lerman07digg}, retweet the URLs their friends post on Twitter~\cite{WuWatts11}, view and favorite friends' photos on Flickr~\cite{Lerman07flickr,Cha09www}, and so on. While friends' activity can be a useful predictor of user's actions, we claim that knowing at least the local structure of the follower graph can enhance the power of this predictor. In other words, while social media users tend to act like their friends, they are more likely to act like their closer friends.

We evaluate this claim on the task of predicting user activity on Digg and Twitter. This task can be stated as follows: given the follower graph and the stories that a user's friends voted for (or retweeted), predict which stories the user votes for (or retweets). Following methodology described in Section~\ref{sec:methodology}, we construct a prediction vector $p$ for a user. The value $p_i$ of the prediction vector represents probability a user's friends voted for the $i^{th}$ URL, weighted by each friend's proximity to the user in the follower graph. To compute precision and recall of prediction, we construct a vector $u$ of URLs the user actually voted for, and compute precision and recall as shown in Algorithm~\ref{alg:predict}. We compare proximity-based prediction to \emph{baseline} that weighs each friend's votes uniformly, without regard to her proximity to the user.

\remove{
\begin{figure}[tbh]
\begin{center}
\includegraphics[width=0.95\linewidth]{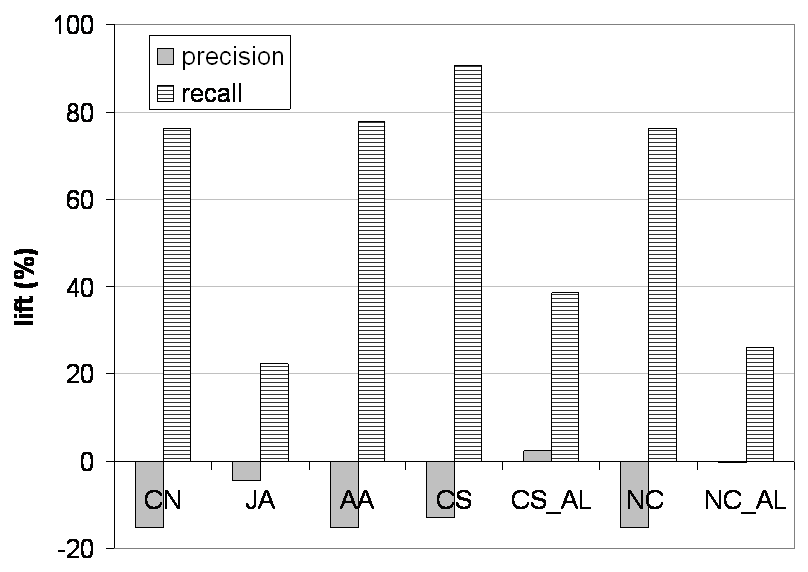}
\end{center}
\caption{Precision and recall lift (\% change over baseline) for predicting votes in the Digg data set.}
\label{fig:digg-prediction}
\end{figure}
}

\begin{table}
  \centering
\caption{Evaluation of predictions by different metrics in the Digg and Twitter data sets. Lift is defined as \% change over baseline.}
\scalebox{0.93}{
\begin{tabular}{@{ }l@{ }|@{ }c@{ }|@{ }c@{ }|@{ }c@{ }|@{ }c@{ }|@{ }c@{ }|@{ }c@{ }|@{ }c@{ }}
  \hline
	&	base	&	\small{CN, NC}	&	\small{JA}	&	\small{AA}	&	\small{CS}	&	\small{CS\_AL}	&	 \small{NC\_AL}	\\
\hline
\multicolumn{8}{c}{\textbf{(a) Digg: all votes}} \\ \hline
precision	&	0.045	&	0.038	&	0.043	&	0.038	&	0.039	&	0.046	&	0.045	\\
recall	&	0.165	&	0.291	&	0.202	&	0.293	&	0.315	&	0.229	&	0.208	\\
 \hline
pr lift \%	&	0	&	-15.3	&	-4.5	&	-15.2	&	-13.0	&	\textbf{2.4}	&	-0.4	\\
re lift \%	&	0	&	{76.3}	&	22.3	&	\textbf{77.8}	&	\textbf{90.7}	&	38.5	&	26.0	\\
 \hline
\multicolumn{8}{c}{\textbf{(b) Digg: pre-promotion votes}}\\ \hline
precision	&	0.032	&	0.027	&	0.033	&	0.027	&	0.028	&	0.039	&	0.034	\\
recall	&	0.172	&	0.248	&	0.174	&	0.250	&	0.272	&	0.195	&	0.174	\\
\hline
pr lift \%	&	0	&	-15.0	&	3.3	&	-14.7	&	-11.1	&	\textbf{22.1}	&	\textbf{7.7}	\\
re lift \%	&	0	&	44.2	&	1.1	&	\textbf{45.5}	&	\textbf{57.9}	&	13.3	&	1.3	\\
\hline
\multicolumn{8}{c}{\textbf{(c) Twitter: all retweets}}\\ \hline
precision	&	0.105	&	0.091	&	0.120	&	0.093	&	0.094	&	0.133	&	0.125	\\
recall	&	0.094	&	0.090	&	0.102	&	0.091	&	0.097	&	0.113	&	0.106	\\
\hline
pr lift \%	&	0	&	-14.1	&	14.1	&	-12.0	&	-10.7	&	\textbf{25.9}	&	\textbf{18.5}	\\
re lift \%	&	0	&	-4.8	&	8.4	&	-3.4	&	2.8	&	\textbf{19.7}	&	\textbf{12.3}	\\
\hline
\end{tabular}
}
\label{tbl:digg}
\end{table}

Voters in the Digg data set voted on more than 3.5K stories. Almost 53K of these voters had at least one friend and were included in the baseline. Of these, we could calculate proximity for about 25K voters. The rest of the voters did not share any common friends or followers with other active users. The average precision and recall values of predicted votes for these users are reported in Table~\ref{tbl:digg}(a).  Although average precision appears low, ranging from 3.8\% to 4.5\%, it is an order of magnitude better than precision of 0.6\% for randomly guessing which stories a user will vote for. We define \emph{lift} as percent-change over baseline. Note that except for conservative attention-limited (CS\_AL), all metrics have worse precision than baseline, although they all have substantially better recall than baseline.

Poor prediction performance appears to contradict our claim that structural proximity helps to predict activity. We can, however, explain this effect by taking into account Digg's user interface. A Digg user can see the activity of her friends, via the friends interface, but she can also see the activity of the entire community via the front page, which shows stories recommended by all users. Digg's front page is the default entry point; therefore, it makes sense that users will often vote for stories they see there, independent of whether they were recommended by friends. These votes may obscure the effect of friends' activities. Before a story is promoted to the front page, however, it can be accessed through the Upcoming stories page, but with tens of thousands of new stories posted to the Upcoming page daily, any individual story will be hard to find. The main driver of votes before promotion is the friend interface, which shows the user stories recommended by friends~\cite{Lerman07ic,Hogg09icwsm}.  Therefore, if we restrict analysis to votes \emph{before} promotion, we should be able to see the network effect of voting. Table~\ref{tbl:digg}(b) reports prediction performance of different metrics for pre-promotion votes only. In this situation, proximity-based prediction results in a substantial lift, as compared to baseline precision, especially for the attention-limited versions of the conservative and non-conservative metrics. Even Jaccard results in a small positive lift, while common neighbors and Adamic-Adar metrics still perform worse than baseline. We conclude that although mass communication via Digg's front page dilutes effect of network-based story recommendation, if we consider network-based communication only, structural proximity can help the prediction task.

In the Twitter data set, almost 542K user retweeted 3.8K URLs. Twitter does not provide an equivalent of Digg's front page for the most retweeted URLs; therefore, URLs generally spread via recommendations by friends. Table~\ref{tbl:digg}(c) compares prediction performance of different proximity metrics. Attention-limited versions of the conservative and non-conservative metrics result in the greatest lift both in precision and recall, up to 25\%. As in the Digg data set, the precision of the common neighbors, Adamic-Adar, and conservative metrics is worse than baseline.

\remove{
\begin{table}
  \centering
  \scalebox{0.8}{
\begin{tabular}{|l|c|c|c|c|c|c|c|}
  \hline
	&	base	&	\small{CN, NC}	&	\small{JA}	&	\small{AA}	&	\small{CS}	&	\small{CS\_AL}	&	 \small{NC\_AL}	\\
\hline
precision	&	0.105	&	0.091	&	0.120	&	0.093	&	0.094	&	0.133	&	0.125	\\
recall	&	0.094	&	0.090	&	0.102	&	0.091	&	0.097	&	0.113	&	0.106	\\
cosine	&	0.189	&	0.209	&	0.222	&	0.212	&	0.201	&	0.209	&	0.216	\\
\hline \\ \hline
pr lift \%	&	0	&	-14.14	&	14.14	&	-11.95	&	-10.72	&	\textbf{25.90}	&	\textbf{18.50}	\\
re lift \%	&	0	&	-4.77	&	8.37	&	-3.39	&	2.75	&	\textbf{19.70}	&	\textbf{12.29}	\\
cos lift \%	&	0	&	10.55	&	\textbf{17.55}	&	12.41	&	6.42	&	11.03	&	\textbf{14.32}	\\
 \hline
\end{tabular}
}
  \caption{Precision and recall lift (percent change over baseline) for predicting retweets in the Twitter data set.}
\label{tbl:twitter}
\end{table}
}

\remove{
\begin{figure}[tbh]
\begin{center}
\includegraphics[width=0.95\linewidth]{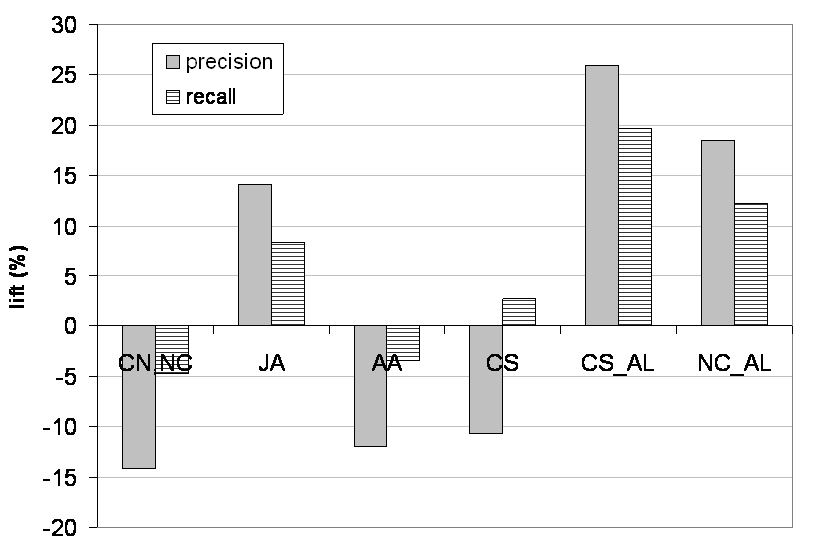}
\end{center}
\caption{Precision and recall lift (\% change over baseline) for predicting retweets in the Twitter data set.}
\label{fig:digg-prediction}
\end{figure}
}

\subsection{Discussion}
Just as in the link prediction task, structural information can help activity prediction task. However, as we show in this paper, the choice of the structural metric matters for prediction performance.
Although non-conservative metric produced the highest correlation between structural proximity and activity, it did not lead to the best prediction performance. In fact, on both Digg and Twitter it gave the worst predictions, compared to the uniform friend recommendation baseline. The non-conservative metrics model epidemic spreading in networks. We know, however, that information spread in social media (at least on Digg) is somewhat different from the spread of epidemics, because probability of becoming ``infected'' with information does not depend on the number of ``infected'' friends~\cite{Versteeg11icwsm}. Results of this paper suggest that attention plays an important role in information spread in social media. Even if we do not yet fully understand this process, we show in this paper the choice of the proximity metric matters. The reason that attention-limited metrics produce the best prediction results is because they more closely describe the dynamic processes taking place in social media than other metrics. This may also help explain link prediction results. The reason Adamic-Adar performed best on the task of predicting future paper co-authorship, probably because of the many metrics studied by Liben-Nowell and Kleinberg, it most closely approximated the nature of interactions between authors, which is probably best modeled by an attention-limited process. On the missing link prediction task in conservative transportation and power grid networks, linear RA metric gave the best results. This makes sense, since the RA metric is an unsymmetrized version of the conservative metric described in this paper. This further underscores the need to consider the nature of the dynamic process when choosing proximity metric for the prediction task.


The values reported in Table~\ref{tbl:digg} represent of precision and recall averaged over all users. Precision values have a heavy-tailed distribution (recall is more uniformly distributed). This means that while for a majority of users precision is almost no better than random guess, for other users relatively high precision can be achieved. It may be possible to automatically distinguish users whose actions we can predict with high confidence from those whose actions are essentially unpredictable. We leave this question for future investigation. Our work also ignores the timing of votes, i.e., whether friends' recommendations came before or after a user's own recommendation. Therefore, we do not distinguish between the effects of homophily and influence~\cite{Shalizi11}. This too will be the subject for future study.

\section{Related Work}
Granovetter~\cite{Granovetter73}  proposed neighborhood overlap as a metric to quantify the strength of a tie, i.e., how intensely and deeply two actors in a social network interact. If $u$ and $v$ have many friends in common, they are more likely to attend the same events and be exposed to the same information, and therefore, interact and act in a similar manner.
A study of a massive mobile phone network established a correlation between social tie strength and neighborhood overlap, or proximity~\cite{Onnela07}. This study measured tie strength by the frequency and duration of phone calls between two people, and it measured proximity by the fraction of common neighbors. Though it established a correlation between proximity and activity, it did not attempt to predict activity. Granovetter's paper is best remembered for the special role he assigned to weak ties in information diffusion. In this paper, we only focus on the role of strong ties in predicting activity.

Activity prediction is similar to the link prediction prediction in that it uses network structure for prediction. However, these problems are fundamentally different, because in link prediction, structural evidence is used to predict structure of the network, while in activity prediction, structural evidence is used to predict user activity, a distinct source of evidence.
Several researchers have studied the link prediction task, in which they used network proximity to identify unobserved or missing links or to predict future links in a network. These studies used a number of metrics, including the number and fraction of common neighbors, Adamic-Adar score~\cite{Liben-Nowell03,linkprediction}, as well as a metric based on resource allocation (RA)~\cite{Zhou09}, and those based on the random walk, such as effective conductance~\cite{Koren06} and escape probability~\cite{Tong07,Tong08}. Although some metrics were shown to perform better than others, no explanation was given for these differences.
On the link prediction task in the co-authorship networks, for example, Adamic-Adar score gave best results~\cite{Liben-Nowell03}, while on the missing link prediction task in power grid and transportation networks, the linear version (RA) of Adamic-Adar performed best~\cite{Zhou09}.  We postulate that the reason RA metric, which is equivalent to our conservative proximity, worked so well is because it captures the conservative nature of interactions in the power grid and transportation networks. We suspect that Adamic-Adar worked best on the link prediction task because of all the metrics tested by Liben-Nowell and Kleinberg, it came closest to capturing the nature of interactions between authors. We suspect that metrics we introduce in this paper will lead to an even better link prediction performance.

Activity and network structure are, of course, not completely independent. Previous studies examined the impact of social ties and network structure on user behavior. Anagnostopoulos et al.~\cite{Anagnostopoulos08} examined user activity on a social media site Flickr and found evidence for social correlations, i.e., they found that user's tagging activity was similar to that of her friends in the network. The goal of that work, however, was to test whether homophily or social influence is responsible for social correlation. Other studies~\cite{Aral09,Choudhury10,Steglich10} have examined the cause of behavior correlation in networks, both online social networks and friendship networks. We do not attempt to explain the source of social correlation and its relationship to network structure, rather we exploit existing correlations to predict activity.

\section{Conclusion}
We introduce activity prediction task for social networks. In this task, information about activity of a user's friends in the social network is used to predict user's activity. We showed that taking into account how close these friends are to the user can help better predict user's activity. In addition to existing proximity metrics, which measure how close nodes are in the network, we defined new metrics that take into account the nature of interactions between nodes in the network. These metrics were inspired by social communication, which is often constrained by finite attention. In other words, the more friends a person has, the less time she can devote to interacting with a specific friend. We explored the performance of these metrics on the task of predicting user activity on social media sites Digg and Twitter. We found that taking into account friends' proximity to the user can improve prediction, and that most gain is achieved by the attention-limited metrics.

This papers opens several new avenues for exploration. Although we did not explore the underlying reasons for correlation between structure and activity, it could be as Granovetter noted, people linked by strong ties act in a similar manner because they belong to the same community. This implies that proximity metrics could be used for community identification task, and that different metrics will lead to different community divisions. We also did not explore the temporal nature of activity, whether user retweets the URL before or after her friend does. In addition, we found evidence that some users' activity may be easier to predict than others, so an interesting question is whether we can automatically determine whose behaviors are more predictable. We leave these questions for future research.


\bibliographystyle{abbrv}
\bibliography{references,../lerman}  

\begin{thebibliography}{10}

\bibitem{adamic03friends}
L.~Adamic and E.~Adar.
\newblock {Friends and neighbors on the Web}.
\newblock {\em Social Networks}, 25(3):211--230, July 2003.

\bibitem{Anagnostopoulos08}
A.~Anagnostopoulos, R.~Kumar, and M.~Mahdian.
\newblock {Influence and correlation in social networks}.
\newblock In {\em Proceeding of the 14th ACM SIGKDD international conference on
  Knowledge discovery and data mining}, KDD '08, pages 7--15, New York, NY,
  USA, 2008. ACM.

\bibitem{Aral09}
S.~Aral, L.~Muchnik, and A.~Sundararajan.
\newblock {Distinguishing influence-based contagion from homophily-driven
  diffusion in dynamic networks}.
\newblock {\em Proceedings of the National Academy of Sciences},
  106(51):21544--21549, Dec. 2009.

\bibitem{Aral11}
S.~Aral and D.~Walker.
\newblock {Creating Social Contagion Through Viral Product Design: A Randomized
  Trial of Peer Influence in Networks}.
\newblock {\em Management Science}, 57(9):1623--1639, Aug. 2011.

\bibitem{Cha09www}
M.~Cha, A.~Mislove, and K.~P. Gummadi.
\newblock {A measurement-driven analysis of information propagation in the
  flickr social network}.
\newblock In {\em Proceedings of the 18th international conference on World
  wide web}, WWW '09, pages 721--730, New York, NY, USA, 2009. ACM.

\bibitem{DGG}
A.~Davis, B.~B. Gardner, and M.~R. Gardner.
\newblock {\em Deep South}.
\newblock The University of Chicago Press, Chicago, 1941.

\bibitem{Choudhury10}
M.~De~Choudhury, H.~Sundaram, A.~John, D.~D. Seligmann, and A.~Kelliher.
\newblock {"Birds of a Feather": Does User Homophily Impact Information
  Diffusion in Social Media?}
\newblock June 2010.

\bibitem{Freeman01}
L.~Freeman.
\newblock {Finding Social Groups: A Meta-Analysis of the Southern Women Data}.

\bibitem{Ghosh11nonconservative}
R.~Ghosh, K.~Lerman, T.~Surachawala, K.~Voevodski, and S.-H. Teng.
\newblock {Non-Conservative} diffusion and its application to social network
  analysis.
\newblock Technical report, University of Southern California, Feb 2011.

\bibitem{Ghosh11snakdd}
R.~Ghosh, T.~Surachawala, and K.~Lerman.
\newblock Entropy-based classification of 'retweeting' activity on twitter.
\newblock In {\em Proceedings of KDD workshop on Social Network Analysis
  (SNA-KDD)}, August 2011.

\bibitem{Granovetter73}
M.~S. Granovetter.
\newblock {The Strength of Weak Ties}.
\newblock {\em American Journal of Sociology}, 78(6):1360--1380, 1973.

\bibitem{Hogg09icwsm}
T.~Hogg and K.~Lerman.
\newblock Stochastic models of user-contributory web sites.
\newblock In {\em Proceedings of 3rd International Conference on Weblogs and
  Social Media (ICWSM)}, May 2009.

\bibitem{Katz53}
L.~Katz.
\newblock {A new status index derived from sociometric analysis}.
\newblock {\em Psychometrika}, 18(1):39--43, Mar. 1953.

\bibitem{Koren06}
Y.~Koren, S.~C. North, and C.~Volinsky.
\newblock {Measuring and extracting proximity graphs in networks}.
\newblock {\em ACM Trans. Knowl. Discov. Data}, 1(3), Dec. 2007.

\bibitem{Lerman07ic}
K.~Lerman.
\newblock Social information processing in social news aggregation.
\newblock {\em IEEE Internet Computing: special issue on Social Search},
  11(6):16--28, 2007.

\bibitem{Lerman07digg}
K.~Lerman.
\newblock Social networks and social information filtering on digg.
\newblock In {\em Proceedings of 1st International Conference on Weblogs and
  Social Media (ICWSM-07)}, 2007.

\bibitem{Lerman10icwsm}
K.~Lerman and R.~Ghosh.
\newblock Information contagion: an empirical study of spread of news on digg
  and twitter social networks.
\newblock In {\em Proceedings of 4th International Conference on Weblogs and
  Social Media (ICWSM)}, May 2010.

\bibitem{Lerman07flickr}
K.~Lerman and L.~Jones.
\newblock Social browsing on flickr.
\newblock In {\em Proceedings of 1st International Conference on Weblogs and
  Social Media (ICWSM-07)}, 2007.

\bibitem{Liben-Nowell03}
D.~Liben-Nowell and J.~Kleinberg.
\newblock {The link-prediction problem for social networks}.
\newblock {\em J. Am. Soc. Inf. Sci.}, 58(7):1019--1031, 2007.

\bibitem{Lotan11}
G.~Lotan, E.~Graeff, M.~Ananny, D.~Gaffney, I.~Pearce, and D.~Boyd.
\newblock {The Revolutions Were Tweeted: Information Flows during the 2011
  Tunisian and Egyptian Revolutions}.
\newblock {\em International Journal of Communications}, 5:1375--1405, 2011.

\bibitem{linkprediction}
L.~L\"{u} and T.~Zhou.
\newblock {Link prediction in complex networks: A survey}.
\newblock {\em Physica A: Statistical Mechanics and its Applications}, Dec.
  2010.

\bibitem{Onnela07}
J.~P. Onnela, J.~Saram\"{a}ki, J.~Hyv\"{o}nen, G.~Szab\'{o}, D.~Lazer,
  K.~Kaski, J.~Kert\'{e}sz, and A.~L. Barab\'{a}si.
\newblock {Structure and tie strengths in mobile communication networks}.
\newblock {\em Proceedings of the National Academy of Sciences},
  104(18):7332--7336, May 2007.

\bibitem{Plangprasopchok11wsdm}
A.~Plangprasopchok, K.~Lerman, and L.~Getoor.
\newblock A probabilistic approach for learning folksonomies from structured
  data.
\newblock In {\em Proceedings of the 4th ACM Web Search and Data Mining
  Conference (WSDM)}, February 2011.

\bibitem{www11-hashtags}
D.~M. Romero, B.~Meeder, and J.~Kleinberg.
\newblock {Differences in the Mechanics of Information Diffusion Across Topics:
  Idioms, Political Hashtags, and Complex Contagion on Twitter}.
\newblock In {\em Proceedings of World Wide Web Conference}, 2011.

\bibitem{Shalizi11}
C.~R. Shalizi and A.~C. Thomas.
\newblock {Homophily and Contagion Are Generically Confounded in Observational
  Social Network Studies}.
\newblock {\em Sociological Methods \& Research}, 40(2):211--239, May 2011.

\bibitem{Versteeg11icwsm}
G.~V. Steeg, R.~Ghosh, and K.~Lerman.
\newblock What stops social epidemics?
\newblock In {\em Proceedings of 5th International Conference on Weblogs and
  Social Media}, 2011.

\bibitem{Steglich10}
C.~Steglich, T.~A.~B. Snijders, and M.~Pearson.
\newblock {Dynamic Networks and Behavior: Separating Selection from Influence}.
\newblock {\em Sociological Methodology}, 2010.

\bibitem{Tong07}
H.~Tong, C.~Faloutsos, and Y.~Koren.
\newblock {Fast direction-aware proximity for graph mining}.
\newblock In {\em KDD '07: Proceedings of the 13th ACM SIGKDD international
  conference on Knowledge discovery and data mining}, pages 747--756, New York,
  NY, USA, 2007. ACM.

\bibitem{Tong08}
H.~Tong, S.~Papadimitriou, P.~S. Yu, and C.~Faloutsos.
\newblock {Proximity Tracking on Time-Evolving Bipartite Graphs}.
\newblock In {\em SIAM Conference on Data Mining (SDM08)}, 2008.

\bibitem{Wu07}
F.~Wu and B.~A. Huberman.
\newblock {Novelty and collective attention}.
\newblock {\em Proceedings of the National Academy of Sciences},
  104(45):17599--17601, Nov. 2007.

\bibitem{WuWatts11}
S.~Wu, J.~M. Hofman, W.~A. Mason, and D.~J. Watts.
\newblock {Who Says What to Whom on Twitter}.
\newblock In {\em Proceedings of World Wide Web Conference (WWW '11)}, 2011.

\bibitem{Zhou09}
T.~Zhou, L.~L\"{u}, and Y.-C. Zhang.
\newblock {Predicting missing links via local information}.
\newblock {\em The European Physical Journal B - Condensed Matter and Complex
  Systems}, 71(4):623--630, Oct. 2009.

\end{thebibliography}
\balancecolumns
\end{document}